\documentclass[tighten, twocolumn]{aastex631}
\usepackage{CJK}

\shorttitle{A Tilted Halo}
\shortauthors{Han et al.}

\graphicspath{{./}{figures/}}

\begin{document}
\begin{CJK*}{UTF8}{gbsn}

\title{A Tilt in the Dark Matter Halo of the Galaxy}

\correspondingauthor{Jiwon Jesse Han}
\email{jesse.han@cfa.harvard.edu}

\author[0000-0002-6800-5778]{Jiwon Jesse Han}
\affiliation{Center for Astrophysics $|$ Harvard \& Smithsonian, 60 Garden Street, Cambridge, MA 02138, USA}

\author[0000-0003-3997-5705]{Rohan P. Naidu}
\affiliation{Center for Astrophysics $|$ Harvard \& Smithsonian, 60 Garden Street, Cambridge, MA 02138, USA}

\author[0000-0002-1590-8551]{Charlie Conroy}
\affiliation{Center for Astrophysics $|$ Harvard \& Smithsonian, 60 Garden Street, Cambridge, MA 02138, USA}

\author[0000-0002-7846-9787]{Ana Bonaca}
\affiliation{The Observatories of the Carnegie Institution for Science, 813 Santa Barbara St., Pasadena, CA 91101, USA}

\author[0000-0002-5177-727X]{Dennis Zaritsky}
\affiliation{Steward Observatory, University of Arizona, 933 North Cherry Avenue, Tucson, AZ 85721-0065, USA}

\author[0000-0003-2352-3202]{Nelson Caldwell}
\affiliation{Center for Astrophysics $|$ Harvard \& Smithsonian, 60 Garden Street, Cambridge, MA 02138, USA}

\author[0000-0002-1617-8917]{Phillip Cargile}
\affiliation{Center for Astrophysics $|$ Harvard \& Smithsonian, 60 Garden Street, Cambridge, MA 02138, USA}

\author[0000-0002-9280-7594]{Benjamin D. Johnson}
\affiliation{Center for Astrophysics $|$ Harvard \& Smithsonian, 60 Garden Street, Cambridge, MA 02138, USA}

\author[0000-0002-0572-8012]{Vedant Chandra}
\affiliation{Center for Astrophysics $|$ Harvard \& Smithsonian, 60 Garden Street, Cambridge, MA 02138, USA}

\author[0000-0003-2573-9832]{Joshua S. Speagle (\begin{CJK*}{UTF8}{gbsn}沈佳士\ignorespacesafterend\end{CJK*})}
\altaffiliation{Banting \& Dunlap Fellow}
\affiliation{David A. Dunlap Department of Astronomy \& Astrophysics, University of Toronto, 50 St. George Street, Toronto ON M5S 3H4, Canada}
\affiliation{Dunlap Institute for Astronomy and Astrophysics, University of Toronto, 50 St George Street, Toronto, ON M5S 3H4, Canada}
\affiliation{Department of Statistical Sciences, University of Toronto, 100 St George St, Toronto, ON M5S 3G3, Canada}

\author[0000-0001-5082-9536]{Yuan-Sen Ting (丁源森)}
\affiliation{Research School of Astronomy \& Astrophysics, Australian National University, Cotter Road, Weston Creek, ACT 2611, Canberra, Australia}
\affiliation{Research School of Computer Science, Australian National University, Acton ACT 2601, Australia}

\author[0000-0002-0721-6715]{Turner Woody}
\affiliation{Center for Astrophysics $|$ Harvard \& Smithsonian, 60 Garden Street, Cambridge, MA 02138, USA}

\begin{abstract}

Recent observations of the stellar halo have uncovered the debris of an ancient merger, Gaia-Sausage-Enceladus, estimated to have occurred $\gtrsim 8\text{ Gyr}$ ago. Follow-up studies have associated GSE with a large-scale tilt in the stellar halo that links two well-known stellar over-densities in diagonally opposing octants of the Galaxy (the Hercules-Aquila Cloud and Virgo Overdensity; HAC and VOD). In this paper, we study the plausibility of such unmixed merger debris persisting over several Gyr in the Galactic halo.  We employ the simulated stellar halo from \citet{n21}, which reproduces several key properties of the merger remnant, including the large-scale tilt. By integrating the orbits of these simulated stellar halo particles, we show that adoption of a spherical halo potential results in rapid phase mixing of the asymmetry.  However, adopting a tilted halo potential preserves the initial asymmetry in the stellar halo for many Gyr.  The asymmetry is preserved even when a realistic growing disk is added to the potential. These results suggest that HAC and VOD are long-lived structures that are associated with GSE and that the dark matter halo of the Galaxy is tilted with respect to the disk and aligned in the direction of HAC-VOD. Such halo-disk misalignment is common in modern cosmological simulations. Lastly, we study the relationship between the local and global stellar halo in light of a tilted global halo comprised of highly radial orbits. We find that the local halo offers a dynamically biased view of the global halo due to its displacement from the Galactic Center.

\end{abstract}

\section{Introduction} \label{sec:intro}

\begin{figure*}[t!]
\centering
    \includegraphics[width=0.85\linewidth]{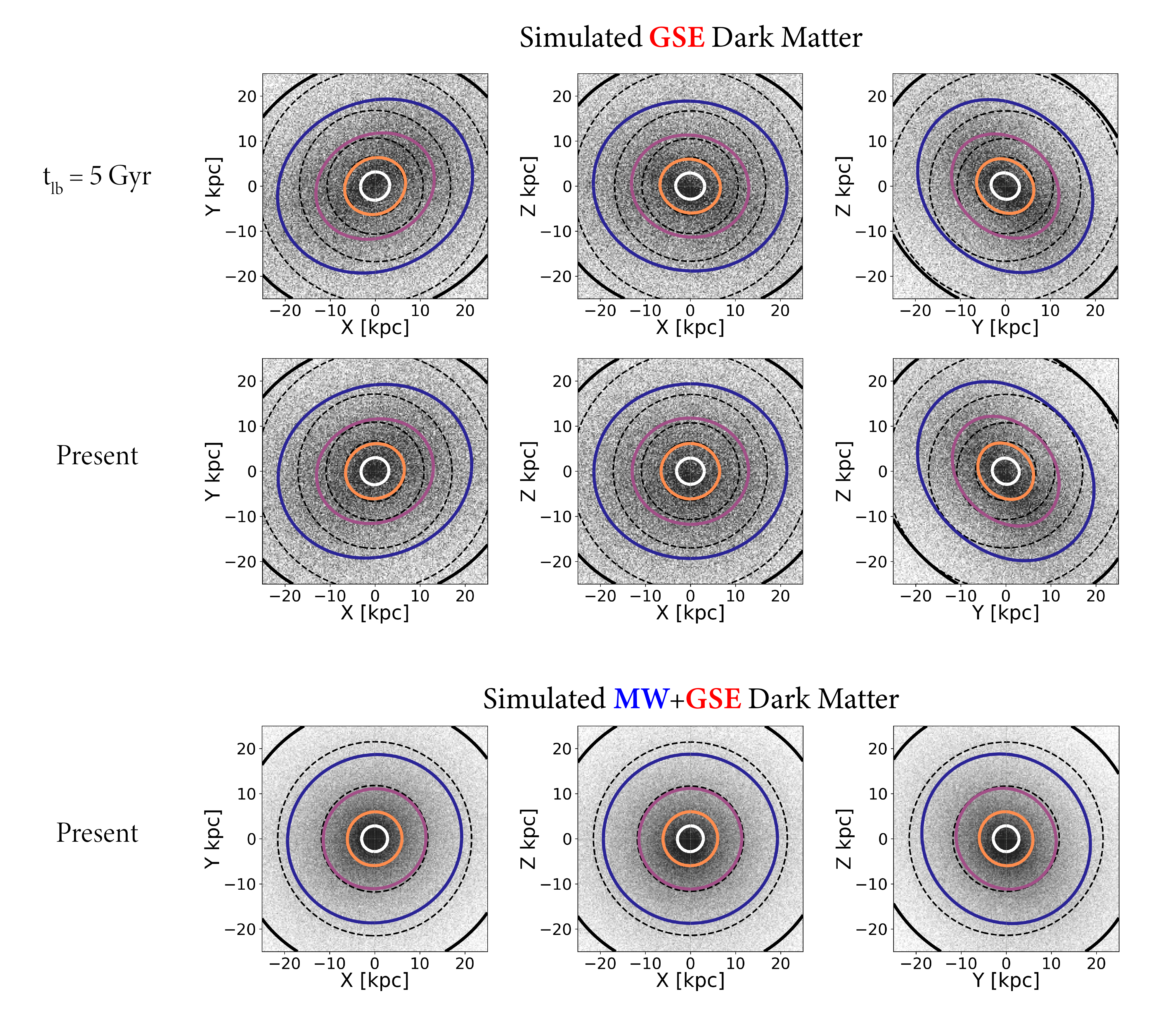}
    \caption{2D histograms of the dark matter (DM) distribution in the fiducial simulation of N21. We adopt a greyscale with asinh stretch that is normalized to include 98\% of the sample. Overplotted in solid lines are equipotential contours in logarithmic spacing, and dashed lines are corresponding contours of a spherical potential to guide the eye. Top and middle panels show DM from the simulated GSE at lookback time $t_{\text{lb}}=$ 5 Gyr and present day, respectively. These panels demonstrate that the DM distribution of GSE is tilted and stable over the last 5 Gyr. Bottom panels show the combined DM distribution of the simulated GSE and MW.}
\label{fig:fig1}
\end{figure*}

Modern stellar surveys (e.g., 2MASS, SDSS, \textit{Gaia}) have carried out a remarkably successful archaeological excavation of our Galaxy's fossil graveyard, the stellar halo. Since the foundational works of \cite{ELS62} and \cite{SZ78}, our picture of the Galactic halo  formation history has evolved to encompass both in-situ and accretion processes, some of which are continuing to the present day as evident in the Sagittarius stream \citep[][]{ibata94, majewski03} and the infalling Magellanic Clouds \citep{Mathewson74,besla07,Zaritsky20}. Further back in the Galaxy's timeline, a striking discovery from the \textit{Gaia} mission suggests that the bulk of the stellar halo formed from a single merger event 8-10 billion years ago. Referred to here as \textit{Gaia}-Sausage-Enceladus (GSE; \citealt{belokurov18}, \citealt{Helmi18}), the debris of this merger comprises the majority of the inner stellar halo (Galactocentric radius $r_{\text{gal}} \lesssim 30 \text{ kpc}$; \citealt{IB19}, \citealt{naidu20}), providing compelling evidence for hierarchical formation of the Galaxy as predicted by the $\Lambda$-CDM standard cosmological model \citep[][]{PS74,blumenthal84,BJ05}.

In light of this discovery, two major stellar overdensities in the halo, the Hercules-Aquila Cloud (HAC, \citealt{HAC}) and the Virgo Overdensity (VOD, \citealt{VOD}), have been interpreted as apocentric pile-ups \citep{Deason18,simion19,n21} of the GSE merger debris. Supporting evidence of this association is that they occupy diagonally opposing octants in the Galaxy (\citealt{IB19}), overlap in orbital properties \citep{simion19}, and exhibit metallicity distributions indistinguishable from that of GSE \citep[][hereafter N21]{n21}. In addition, Han et al. (in prep) measure the global shape and orientation of the stellar halo using a sample of giants from the H3 survey \citep{H3} equipped with 6D phase space, metallicity, and $[\alpha/\text{Fe}]$ abundance measurements beyond distances $d \sim 50 \text{ kpc}$. Using a chemodynamical selection to identify GSE giants, they find that the diffuse halo substructure is best fit by a prolate spheroid, tilted off the Galactic plane pointing towards VOD in the North, and HAC in the South. This is consistent with the distribution of \textit{Gaia} DR2 RR Lyrae in the halo reported by \cite{IB19}.

\begin{figure*}[t!]
    \centering
    \includegraphics[width=0.85\linewidth]{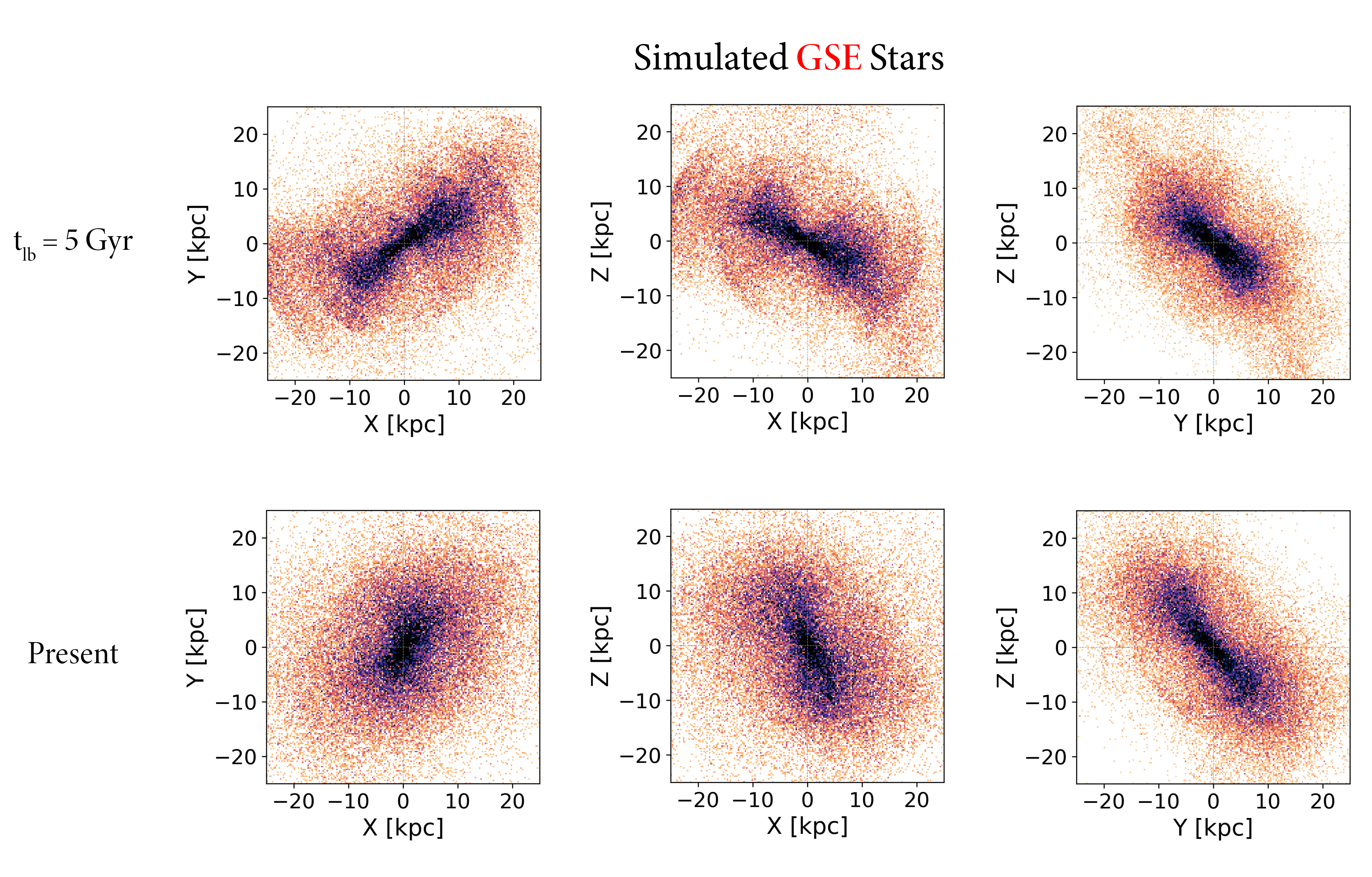}
    \caption{2D histograms of the simulated GSE stellar distribution in N21 in three projections, adopting an asinh stretch that is normalized to include 98\% of the sample. All following figures with 2D histograms adopt the same color scheme. The top and bottom panels show the stellar distribution at lookback time $t_{\text{lb}}=$ 5 Gyr and present day, respectively. These panels show that the tilt in the stellar distribution of the simulated GSE persists over 5 Gyr.}
    \label{fig:fig2}
\end{figure*}

Parallel to these efforts constraining the spatial extent of GSE, the timing of the merger has also been investigated. Stellar ages estimated from various techniques have been used to infer a timeline of the merger \citep[][]{gallart19, bonaca20, belokurov20, chaplin20, borre21, Grunblatt21, montalban21}, building a consensus that the merger completed $8-10\text{ Gyr}$ ago. This merger timeline is also supported by the kinematics of in-situ halo stars, which are consistent with being heated from the thick (high-$\alpha$) disk due to GSE \citep{Zolotov09,Purcell10,bonaca17}. In simulations, \cite{fattahi19} identify halos in the Auriga project \citep{auriga} that host an accreted, highly radial stellar halo to find that these halos are preferentially created from a single merger with a dwarf galaxy $6-10\text{ Gyr}$ ago. Furthermore, \cite{grand18} and \cite{mackereth18} demonstrate using Auriga and EAGLE simulations that the bimodality of the $[\alpha/\text{Fe}]$-$[\text{Fe/H}]$ sequence in the Galactic disk could be the consequence of an early ($z>1$) star formation event triggered by a gas-rich merger. This is once again consistent with the observationally inferred timing of the GSE merger.

\begin{figure*}[t!]
\centering
    \includegraphics[width=0.95\linewidth]{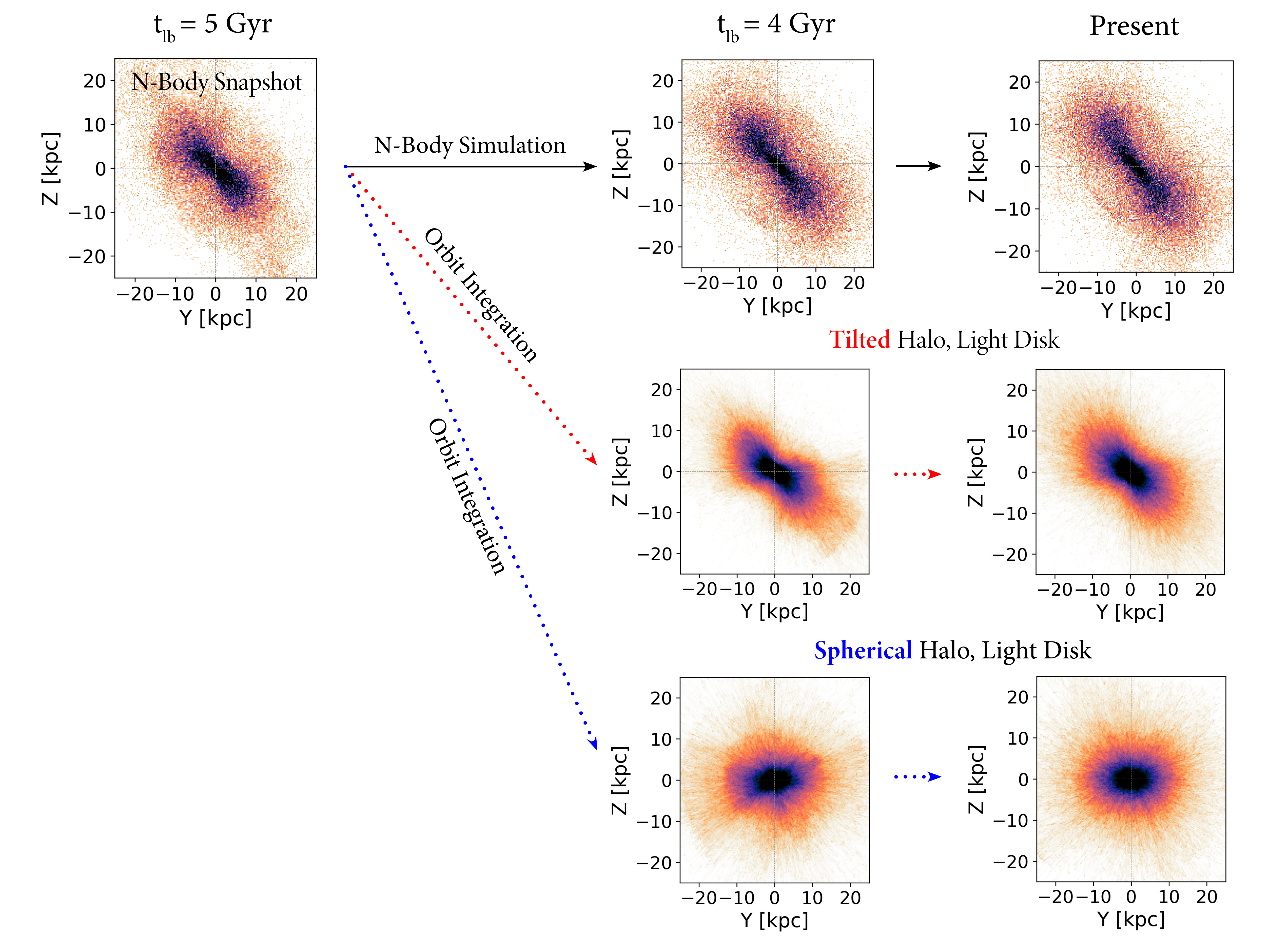}
    \caption{Stellar halo distribution in the Galactocentric YZ projection at three representative time steps ($t_{lb}=\text{5 Gyr}$, $t_{lb}=\text{4 Gyr}$, and present day). Top row shows N-body snapshots from N21, and middle and bottom rows show orbital integrations in a tilted halo and a spherical halo.}
    \label{fig:fig3}
\end{figure*}

While cosmological simulations have been helpful to constrain general properties of the merger, a tailored simulation is required to study the initial conditions of the merger and its subsequent evolution in detail. To this end, N21 present a grid of $\sim500$ N-body simulations from which they identify a fiducial simulation that closely resembles GSE based on H3 data \citep{H3}. This simulation is a 2.5:1 total mass ratio merger of the Milky Way at $z=2$ with a dwarf galaxy on an inclined, retrograde orbit. The resulting z-component of the angular momentum $L_z$ and galactocentric distance $r_{\text{gal}}$ distributions of the merger match those of the GSE sample from the H3 survey \citep{naidu20}. Remarkably, the simulation also reproduces HAC/VOD-like stellar overdensities in the correct galactocentric locations, which was not a condition used to select the GSE analog. Furthermore, these HAC/VOD analogs persist over several Gyr, which is consistent with the observationally inferred timeline of the GSE merger. Lastly, the dark matter contribution from GSE is $\sim20$\% within $r_{\text{gal}}<30\text{ kpc}$ in this simulation, which is consistent with hydrodynamical simulations of GSE-like major mergers in MW-like galaxies \citep{fattahi19,dillamore21}.

Thus, a coherent picture emerges: the stellar halo out to $30\text{ kpc}$ is dominated by debris from an ancient ($\sim8-10\text{ Gyr}$) radial merger and is globally asymmetric, bookended by the two diagonal apocenters at VOD (above the disk) and HAC (below the disk).

Alternatives to this picture have been proposed. For example, \cite{balbinot21} integrate the orbits of a local (heliocentric distance $d < 2.5 \text{ kpc}$) \textit{Gaia} EDR3 6D sample to predict the global distribution of halo substructures. After integrating GSE stars over $8\text{ Gyr}$, they report a symmetric distribution of orbits on the sky. From this, they argue that the current distribution of GSE cannot be globally asymmetric, and thus the observed stellar halo asymmetries in \cite{IB19} could be an artifact of observational biases arising from the \textit{Gaia} scanning pattern. Also based on orbital integration, \cite{donlon19, donlon20,donlon21} conclude that the radial debris observed in the stellar halo must have been deposited recently ($\sim 2 \text{ Gyr}$) in order to explain the spatial coherence of VOD. According to these authors, a large-scale asymmetry in the stellar halo phase mixes too quickly to be compatible with the proposed merger epoch of GSE.

Both of these studies assume a spherical dark matter (DM) halo to integrate orbits of halo stars. Indeed, that is common practice: although observational constraints on the shape of the DM halo are highly uncertain \citep[e.g.,][]{Read14, BH16, desalas21}, a spherical DM halo serves as the default model for the purpose of orbital integration. More sophisticated models still consider an oblate \citep{vc13,hattori21}, prolate \citep{bowden16}, or triaxial halos \citep{lm10} that are aligned with the Galactic disk. However, there are well grounded motivations to consider a DM halo that is tilted with respect to the disk. First, observations of the stellar halo suggest a large-scale tilt on the relevant scales of $r_{\text{gal}} \sim30\text{ kpc}$, hinting that the DM halo may embody a similar tilt. Second, cosmological simulations show that it is common for DM halos in MW-like galaxies to exhibit non-spherical shapes that are misaligned with the disk \citep[e.g.,][]{prada19,emami21,dillamore21} for a variety of reasons linked to their accretion history. In the Galaxy, such a disk-halo misalignment has been proposed to explain the stability of the Sagittarius stream \citep{debattista13} or the co-planar orbits of Galaxy's satellites \citep{shao21}.

Inspired by these clues, in this paper we study the evolution of radial stellar debris in a spherical halo and a tilted halo, and thus the plausibility of unmixed merger debris such as HAC/VOD persisting over several Gyr in the Galactic halo. The rest of the paper is organized as follows. In section \ref{sec:galpot} we define the Galactic potentials of a spherical and tilted DM halo. In section \ref{sec:orbits} we present the results of integrating stellar halo orbits in these potentials. Finally, in section \ref{sec:discuss} we interpret these results in the context of the Galaxy.

\begin{figure*}[t!]
\centering
    \includegraphics[width=0.85\linewidth]{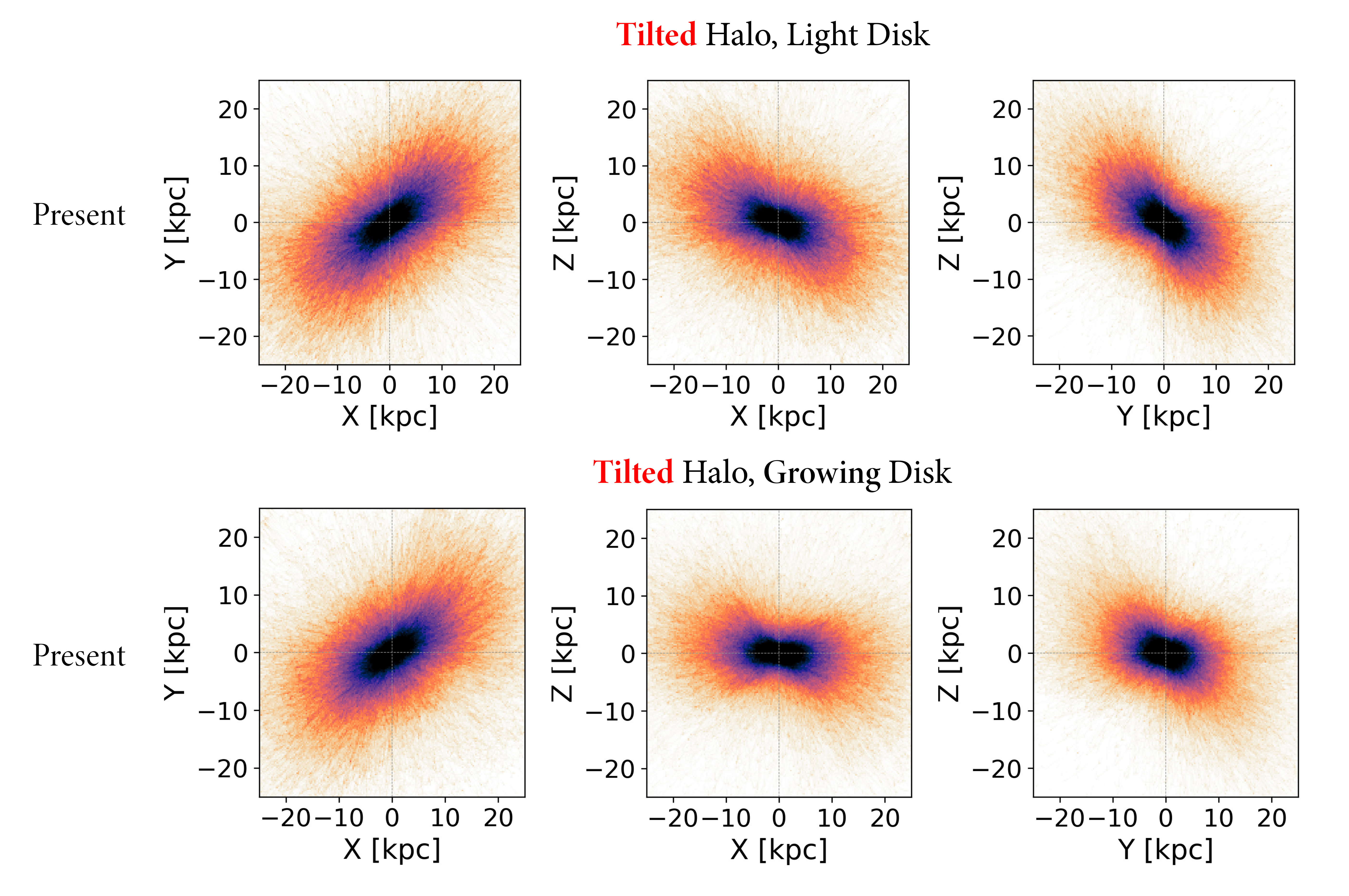}
    \caption{Present day distribution of simulated GSE stars evolved in a tilted, light disk potential (top panels) and a tilted, growing disk potential (bottom panels) in three projections. In the latter case, the disk grows up to six times heavier than its initial mass. Despite the addition of this massive disk to the potential, the tilt of GSE persists over 5 Gyr.}
    \label{fig:fig4}
\end{figure*}
\vspace{10mm}
\section{Galactic Potential} \label{sec:galpot}

In this section, we construct the Galactic potentials used to study the evolution of radial stellar debris in the halo.

We begin by defining the spherical and tilted DM halo models. The former is the canonical picture of the Galactic halo and has been explored to great extent. For this study, we choose the DM halo described by \cite{Bovy15}, following an NFW profile \citep{NFW} with scale radius $r_s=16$ kpc and virial mass $M_{\text{vir}}=8\times10^{11} M_\odot$. The purpose of a tilted halo model in this study is to explore the evolution of large scale stellar halo asymmetries like HAC/VOD, which N21 successfully reproduces. Thus, we adopt the N21 DM distribution as the fiducial tilted halo model.

Figure \ref{fig:fig1} shows the DM distribution in N21. Each panel is a 2D histogram of the positions of DM particles, adopting a greyscale that is normalized to include 98\% of the sample. Overplotted in solid lines are equipotential contours on a logarithmic scale, and dotted lines mark corresponding contours of the spherical halo from \cite{Bovy15} to guide the eye. We show the simulated GSE DM distribution at $t_{\text{lb}}=\text{5 Gyr}$ in the top panels, and at present day in the middle panels. These two rows demonstrate that the GSE DM distribution is tilted and stable over the last 5 Gyr. The bottom panels show the combined DM distribution of the GSE and MW analogs at present day. By fitting a triaxial ellipsoid to these distributions, we find that the GSE DM is triaxial (axis ratio 1:0.9:0.57) and the MW DM is oblate (axis ratio 1:1:0.9), resulting in an overall oblate (axis ratio 1:0.96:0.8) DM distribution that is tilted $\sim30^{\circ}$ above the Galactic disk towards the VOD in the North and HAC in the South.

To study the evolution of stellar orbits in these halos, we now calculate the Galactic potential by combining models of the halo, disk, and bulge.

The potential generated by the spherical halo can be analytically derived from the NFW profile \citep[e.g.,][]{lokas01}. Meanwhile, the tilted DM halo does not follow any particular analytical density function, so we employ the self-consistent field method (SCF; \citealt{Hernquist92}, \citealt{Lowing11}) implemented in the python package \texttt{gala} \citep{gala} to approximate the potential. The advantage of this framework is its flexibility to approximate non-spherical distributions to the desired level of accuracy. After analyzing the level of variance in the terms of the SCF expansion, we include expansion terms up to $n=20$ and $l=4$ to sufficiently reproduce the potential while keeping computational costs manageable. The amplitude of the SCF expansion is determined by matching the energies of orbits near the solar neighborhood to that derived from the spherical potential.

The Galactic disk is another important component of the Galactic potential. While N21 self-consistently describes the dynamics of the merging Galaxy at $z=2$, its disk model mimics the thick disk of the Galaxy and stays at a fixed mass. In the real Galaxy, the thin (low-$\alpha$) disk is $\sim6$ times more massive than the thick (high-$\alpha$) disk at the present day \citep{BH16}. Observations suggest that the bulk of the thin disk grew steadily after $z\sim1-2$ \citep[e.g.,][]{haywood13, fantin19,bonaca20}, following a rapid star formation and quenching  history of the thick disk at $z\sim2$. Over 5 Gyr, an average GSE star can cross the disk up to $\sim10$ times, so the significant growth of the thin disk would clearly impact the dynamics of halo stars. Thus, to examine the effect of the growing thin disk, we consider two models. We first consider a static thick disk with fixed mass $6\times10^{9} M_\odot$, scale length $2\text{ kpc}$, and scale height $900\text{ pc}$. This model is analogous to the disk in N21 and the present day thick disk of the Galaxy. We then consider the thick disk coupled with a linearly growing thin-disk with scale length $2.6\text{ kpc}$ and scale height $300\text{ pc}$ as described in \cite{BH16}. We increase the mass of the thin disk at every Myr, starting from $0$ and ending at $3.5\times10^{10} M_\odot$. This model aims to capture the growth of the thin disk in the Galaxy. Throughout this paper, we refer to the static thick disk model as the ``light disk" and the thick and thin disk model as the ``growing disk."

Finally, for the bulge, we adopt a Hernquist profile \citep{Hernquist90} with mass of $1.4\times10^{10} M_\odot$ and scale length of $1.5\text{ kpc}$ \citep[][]{BH16}. 

We thus consider four Galactic potentials in this study: a tilted/spherical halo with a light disk, and a tilted/spherical halo with a growing disk. In these potentials, we integrate the orbits of stars from the $t_{\text{lb}}=\text{5 Gyr}$ snapshot in N21 to the present day. Figure \ref{fig:fig2} shows the N-body snapshots of the stellar halo at these two epochs. We integrate orbits using the \cite{DP80} integrator at $1 \text{ Myr}$ timestep, implemented in the python package \texttt{gala} \citep{gala}. 

\section{Orbits in the Stellar Halo} \label{sec:orbits}

Here we present the results from integrating stellar halo orbits in the four Galactic potentials. We outline results separately for the global stellar halo and the local (solar neighborhood) halo, the latter being relevant to studies that utilize local 6D data (e.g., the commonly used \textit{Gaia} DR2 and EDR3 RVS samples).
\begin{figure}[t!]
\centering
    \includegraphics[width=\linewidth]{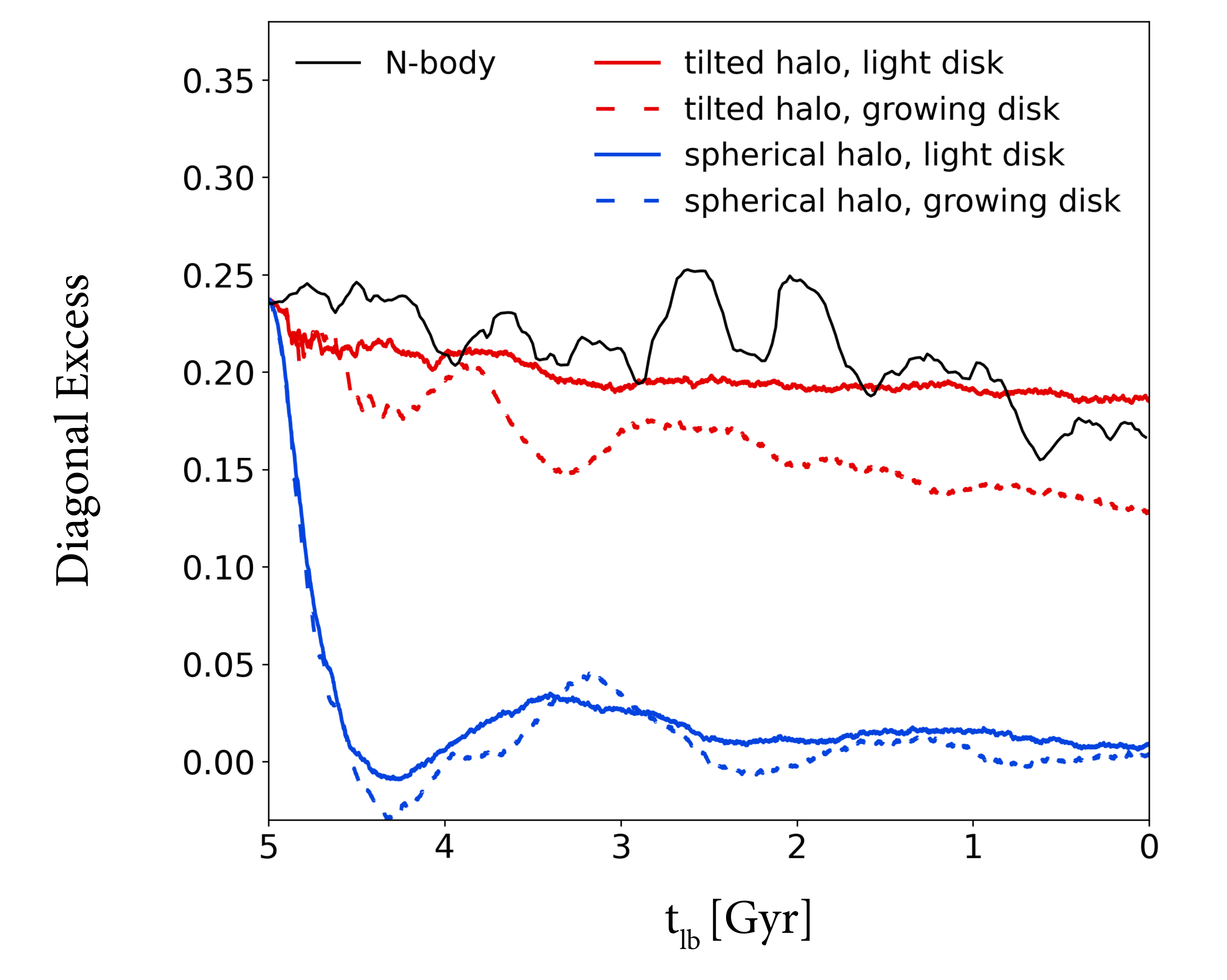}
    \caption{Excess fraction of stars in diagonally opposing octants of the Galaxy as a function of time, defined in equation \ref{eq1}. This diagonal excess is an indicator of the degree to which orbits are spatially mixed. In the N-body and tilted halo models, the initial asymmetry persists beyond $5\text{ Gyr}$, while in the spherical halo models the asymmetry is erased by the first Gyr.}
    \label{fig:fig5}
\end{figure}

\begin{figure}[t!]
\centering
    \includegraphics[width=\linewidth]{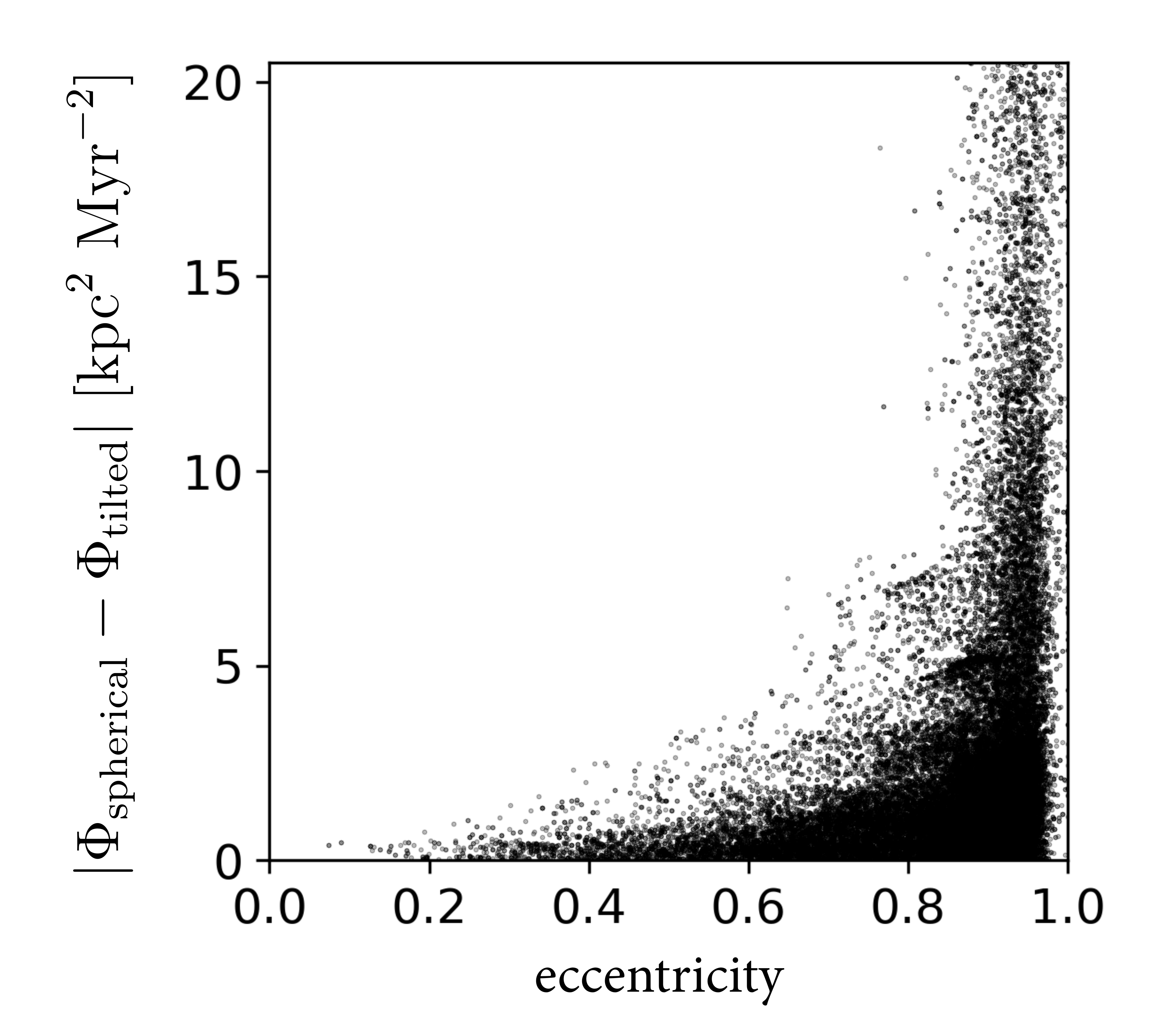}
    \caption{Relation between eccentricity and $|\Phi_{\text{spherical}}-\Phi_{\text{tilted}}|$, where the latter is defined to be the difference in the potential energy that a star experiences over 600 Myr in a tilted potential and a spherical potential as defined in equation \ref{eq2}. There is a clear positive correlation in eccentricity and $|\Phi_{\text{spherical}}-\Phi_{\text{tilted}}|$, demonstrating that stars on eccentric (radial) orbits are more sensitive to its host potential compared to stars on disky orbits.}
    \label{fig:fig6}
\end{figure}

\begin{figure*}[htbp]
\centering
    \includegraphics[width=0.85\linewidth]{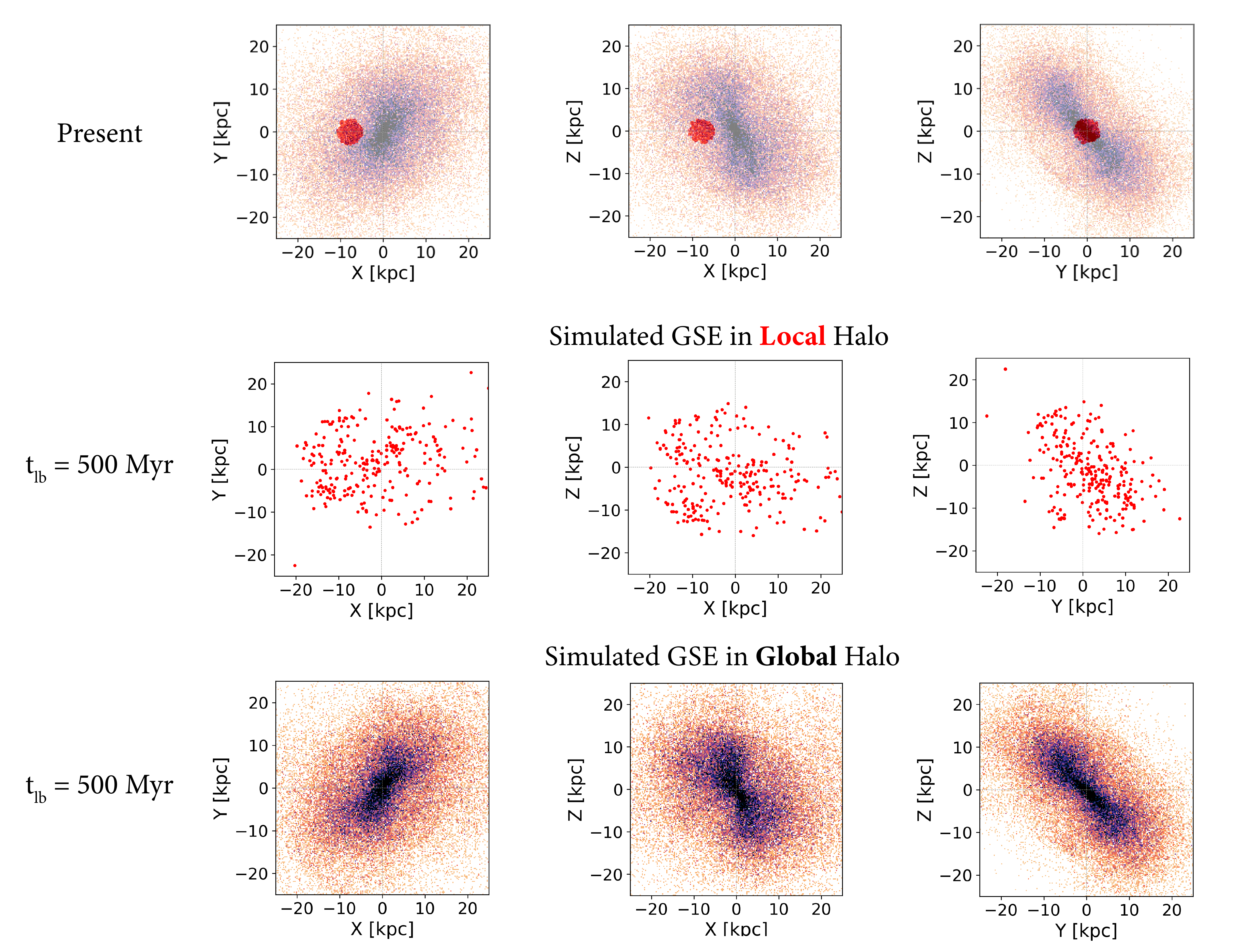}
    \caption{Origin of local halo stars in the N21 simulation. The top row shows the present-day stellar distribution, with local halo stars marked in red. The middle row shows the distribution of local halo stars $500\text{ Myr}$ ago, and the bottom row shows the global halo $500\text{ Myr}$ ago. In the XY and XZ projections, the $t_{lb}=500\text{ Myr}$ distributions of the local and global halo are notably different. This demonstrates that the orbits of local halo stars do not directly trace the spatial structure of the global halo. Meanwhile, in the YZ plane, the local and global halo stars occupy similar distributions, suggesting that in some projections the local halo orbits can reflect the spatial structure of the global halo.}
    \label{fig:fig7}
\end{figure*}

\subsection{Global Halo}\label{sec:results-1}
We show the evolution of the global distribution of halo stars in Figure \ref{fig:fig3}. The leftmost panel shows the $t_{\text{lb}}=\text{5 Gyr}$ snapshot in N21, followed by two columns showing the $t_{\text{lb}}=\text{4 Gyr}$ and present day distributions. In the top row we show the N-body simulations. In the middle and bottom rows, we show the positions of stars derived from orbital integrations at their respective $t_{\text{lb}}$ and the preceding 100 Myr. We choose to display the Galactic YZ plane, and note that the XZ and XY planes show the same trend.

Figure \ref{fig:fig3} showcases how the shape of the halo can greatly influence the evolution of stellar orbits, and we highlight two important features. First, the N-body snapshots and tilted halo orbits closely resemble each other, demonstrating the sufficiency of the SCF approximation of the tilted halo potential. Second, the spherical halo erases the tilt in the stellar halo within the first Gyr, whereas the tilted halo preserves the tilt over $5\text{ Gyr}$. This is somewhat to be expected, since spherical potentials host rosette orbits that can fill the entire available space, while triaxial potentials yield more complex orbits.

Next, Figure \ref{fig:fig4} shows the effect of a growing disk on stellar halo orbits. We show the present day stellar distribution from orbital integration in a tilted halo and light disk in the top panels, and tilted halo and growing disk in the bottom panels. Most notably, the tilt in the stellar halo with respect to the disk is preserved in both potentials. As expected, orbits in the XZ and YZ planes are somewhat concentrated onto the disk, but the spatial asymmetry is still visibly clear. This indicates that the growth of the thin disk in the Galaxy would not erase HAC/VOD like overdensities in the halo.

Figure \ref{fig:fig5} further quantifies the degree to which the stellar halo is spatially mixed. We compute the ``diagonal excess'' of the stellar halo at each time step, defined as
\begin{equation}\label{eq1}
    \text{Diagonal Excess} \equiv \frac{ f_{+\text{X} +\text{Y} -\text{Z}}+f_{-\text{X} -\text{Y} +\text{Z}}}{2} - \frac{ f_{\text{other}}}{6}
\end{equation}
where $f_{+\text{X} +\text{Y} -\text{Z}}$ and $f_{-\text{X} -\text{Y} +\text{Z}}$ are the fraction of stars contained in two diagonal octants (respectively containing the bulk of HAC/VOD in the Galaxy) compared to the fraction of stars in the remaining six octants ($f_{\text{other}}$).  For example, a stellar halo that is tilted and aligned with HAC/VOD will show a positive diagonal excess, while a spherical stellar halo will show a diagonal excess of zero. Thus, if the GSE debris becomes fully mixed, the diagonal excess will converge to zero, while if a large scale asymmetry persists, the diagonal excess will be positive. Figure \ref{fig:fig5} clearly demonstrates this trend. While the N-body simulation (black line) and tilted halo models (red lines) show mixing timescales greater than $5\text{ Gyr}$, the spherical halo (blue lines) fully mixes its orbits by the first Gyr. The growing disk (dashed lines) acts to catalyze the spatial mixing of orbits, but not strongly enough to erase global stellar halo asymmetries evolving in a tilted DM halo. We note that when we integrate orbits in a spherical halo only (excluding the disk and the bulge), there is a remaining diagonal excess of $\sim 0.05$ by 5 Gyr. This demonstrates that while the sphericity of the potential is the driving factor in preserving or destroying the large scale asymmetry of GSE, the disk also plays a role in breaking the spherical symmetry to catalyze the mixing of orbits.

Lastly, Figure \ref{fig:fig6} provides insight into why GSE could be a sensitive probe of the underlying potential. We plot the orbital eccentricity of each star in N21 against $|\Phi_{\text{spherical}}-\Phi_{\text{tilted}}|$, where the latter is defined as

\begin{equation}\label{eq2}
    |\Phi_{\text{spherical}}-\Phi_{\text{tilted}}| \equiv 
    \left | \sum^{600\text{Myr}}_{i=0\text{Myr}}\phi_{\text{spherical},i}-\phi_{\text{tilted},i} \right |
\end{equation}
where $\phi_{\text{spherical,i}}$ and  $\phi_{\text{tilted,i}}$ are the spherical/tilted potentials evaluated at a star's location at $i$ Myr. The 600 Myr summation timescale is chosen such that most stars would have had enough time for at least one pericenter passage. The eccentricity is calculated in the tilted potential, and we make note that the result does not change when eccentricity is calculated in a spherical potential. There is a strong positive correlation in eccentricity and $|\Phi_{\text{spherical}}-\Phi_{\text{tilted}}|$, and the stars on disky orbits show $|\Phi_{\text{spherical}}-\Phi_{\text{tilted}}|$ converging to zero. This correlation indicates that eccentric orbits that probe the full range of radii in the halo are more sensitive to the underlying potential than disky orbits. Since GSE stars are predominantly on highly radial/eccentric orbits, it is a sensitive tracer of the shape of the DM halo that its progenitor deposited in the Galaxy.

\subsection{Local Halo}

We define the local (solar neighborhood) halo to be a sphere of $d\lesssim3\text{ kpc}$ centered on the Sun. The local halo sample has been historically important as it contains the most complete 6D information of stars. Numerous studies have extrapolated the motions of local halo stars in order to understand the global halo \citep[e.g.,][]{carollo10}, and in this section we assess how well the local halo sample represents properties of the global halo.

In Figure \ref{fig:fig7} we identify local halo stars in N21 that satisfy $d<3\text{ kpc}$ at the present day (top row, marked in red), and track where they were 500 Myr ago (middle bottom rows). If the local halo is an unbiased sample of the global halo, then we expect these two distributions to look similar. However, in the XY and XZ projections, the $t_{lb}=500\text{ Myr}$ distribution of the global and local halo are notably different. This suggests that the orbits of local halo stars do not reflect the spatial distribution of the global halo, but are rather on more spatially mixed orbits. This is not surprising, since the solar location is off-axis from the tilt of the global halo and off-centered from the Galactic Center.

Examining the YZ projection in Figure \ref{fig:fig7}, we see that the bottom two panels are similar to each other. This suggests that the local halo stars retain memory of the global halo in the YZ plane, since the solar location on the YZ plane is not off-centered with respect to the Galactic Center. The exact relationship between the local and global halo is a complicated product of the spatial/dynamical distribution of the global halo and the solar position, and cannot be decoupled from one's assumptions about the global halo.

\section{Discussion}\label{sec:discuss}

In this study we have investigated the evolution of stellar debris from an inclined merger remnant in the Galactic halo. Starting with a simulated stellar halo from \cite{n21}, we integrated stellar orbits in tilted and spherical DM halos with both light and growing disks to present day. After $5\text{ Gyr}$, orbits in a spherical halo are fully phase mixed, while orbits in a tilted halo preserve a large-scale tilt even in the influence of a growing (massive) disk. We further quantified stellar halo asymmetries by calculating the excess fraction of stars in the two octants containing HAC/VOD, from which we evaluated the timescales of spatial mixing in the Galactic halo. Finally, we examined the orbits of local halo stars, and showed that these orbits to not trace the spatial distribution of the global halo.

We now interpret these results in the context of the Galactic halo. The interpretation of HAC/VOD as apocenter pileups from GSE is supported by compelling evidence from spatial, dynamical, and chemical considerations.  A key issue is whether one would expect HAC and VOD to be long-lived structures, in light of the evidence that the GSE merger occurred $\sim 8-10\text{ Gyr}$ ago.  If the DM halo of the Galaxy is spherical (or in general is aligned with the disk), then any large-scale asymmetries traced by radial merger debris will be quickly erased.  In this case it would be difficult to understand how HAC and VOD could be associated with an ancient merger.  However, a tilt in the DM halo, as we have explored in this work, enables these over-densities to persist for many Gyr.  The evidence supporting an association between HAC/VOD and GSE, combined with the estimated age of the merger, leads us to conclude that the DM halo of the Galaxy is likely tilted with respect to the disk in the direction of the HAC and VOD over-densities.

In a globally tilted halo, the local halo occupies a spatially and dynamically biased location due to its offset from the Galactic Center. In particular, any sample of stars on radial orbits in the local halo will preferentially contain orbits that have been phase mixed in the Galaxy. In certain Galactocentric projections (e.g., YZ) where the solar neighborhood is not offset from the Galactic Center, these orbits can still retain information about the global halo. In other projections, most of this information is lost. Inferring properties of the global halo by extrapolating the local halo must therefore be carried out with caution.

The non-trivial shape of DM halos has been the topic of many recent studies using cosmological simulations. For example, \cite{emami21} report that only 32\% of Milky Way-like galaxies in the Illustris TNG50 simulation \citep{pillepich19,nelson19} show simple DM halos, and the rest of the sample exhibit either twisted (gradually rotating) or stretched (abruptly rotating) halos. \cite{prada19} find that 20\% of Milky-Way like galaxies in the Auriga project \citep{auriga,monachesi19} have twisted halos, while \cite{dillamore21} find that most galaxies in the \texttt{ARTEMIS} simulations \citep{font20} that show GSE-like features exhibit a global change in orientation of their DM halo. In \texttt{EAGLE} simulations \citep{schaye15,crain15}, \cite{shao21} find that the DM halo of galaxies are well aligned with the common orbital plane of their satellite galaxies. They notice that the common orbital plane of the Galaxy's satellites is misaligned with the disk, from which they argue that the DM halo of the Galaxy must be twisted. In isolated simulations, \cite{debattista13} predicted a halo-disk misalignment in the Galaxy based on models of the Sagittarius stream. Thus, the halo-disk misalignment that we find in this study is consistent with theoretical expectations.

    An important caveat of this study is that we do not consider the mutual interaction between the halo and its environment: the disk and the satellites of the Galaxy. For example, the disk will pull the tilt of the halo closer to the plane, while the halo may induce a warp in the disk at large radii. This could mean that the initial tilt of the halo has to be larger than the current day configuration to explain the spatial distribution of HAC/VOD. We note that GSE contributes only $\sim10$\% of the dark matter within $r_{\text{gal}}<10\text{ kpc}$ in the N21 simulation, so where the disk gravity is dominant the contribution from GSE is smaller than in the outer halo. Among the satellites of the Galaxy, the Large Magellanic Cloud (LMC) has a significant influence on the global shape of the halo \citep[e.g.,][]{GC19, erkal20, conroy21, PP21}. However, this effect is most prominent at larger distances ($r_{\text{gal}}>50\text{ kpc}$) than what we consider here, and within $r_{\text{gal}}<30\text{ kpc}$ the influence of the LMC is likely to be much smaller compared to GSE (though see \citealt{lucchini21} for an alternative scenario). Smaller satellites also contribute to the steady growth in the mass of the halo, but this accretion again occurs mainly in the outer halo while the inner halo is expected to remain quiescent after $t_{\text{lb}}\sim8\text{ Gyr}$ \citep[e.g.,][]{BJ05,font06,dl08,cooper10,pillepich14,BH16}.

Future work will address these issues. Additional observations are necessary to constrain the precise locations and extent of HAC and VOD and the overall tilt of the stellar halo, which will inform the configuration of the DM halo. In light of ongoing and upcoming spectroscopic surveys such as H3, DESI, SDSS-V, WEAVE, and 4MOST, these are timely endeavors.

\bibliography{bibi}{}
\bibliographystyle{aasjournal}

\end{CJK*}
\end{document}